\documentclass[11pt]{article}
\usepackage{amsmath,amssymb,cite}

\makeatletter
\@addtoreset{equation}{section}
\makeatother

\def\ben{\begin{equation}}
\def\een{\end{equation}}

\let\a=\alpha    
    
 \let\m=\mu \let\n=\nu  \let\p=\pi

\let\C=\Chi

\def\nn{\nonumber} \def\bd{\begin{document}} \def\ed{\end{document}}
\def\ds{\documentstyle} \let\fr=\frac \let\bl=\bigl \let\br=\bigr
\let\Br=\Bigr \let\Bl=\Bigl
\let\bm=\bibitem
\let\na=\nabla
\let\pa=\partial \let\ov=\overline
\newcommand{\be}{\begin{equation}}
\newcommand{\ee}{\end{equation}}
\def\ba{\begin{array}}
\def\ea{\end{array}}
\def\ft#1#2{{\textstyle{\frac{\scriptstyle #1}{\scriptstyle #2} } }}
\def\fft#1#2{{\frac{#1}{#2}}}
\def\del{\partial}
\def\vp{\varphi}
\def\sst#1{{\scriptscriptstyle #1}}
\def\oneone{\rlap 1\mkern4mu{\rm l}}
\def\td{\tilde}
\def\wtd{\widetilde}
\def\ie{{\it i.e.\ }}
\def\dalemb#1#2{{\vbox{\hrule height .#2pt
        \hbox{\vrule width.#2pt height#1pt \kern#1pt
                \vrule width.#2pt}
        \hrule height.#2pt}}}
\def\square{\mathord{\dalemb{6.8}{7}\hbox{\hskip1pt}}}
\newcommand{\ho}[1]{$\, ^{#1}$}
\newcommand{\hoch}[1]{$\, ^{#1}$}
\newcommand{\bea}{\begin{eqnarray}}
\newcommand{\eea}{\end{eqnarray}}
\newcommand{\ra}{\rightarrow}
\newcommand{\lra}{\longrightarrow}
\newcommand{\Lra}{\Leftrightarrow}
\newcommand{\bp}{\tilde \beta^\prime}
\newcommand{\tr}{{\rm tr} }
\newcommand{\Tr}{{\rm Tr} }
\def\0{{\sst{(0)}}}
\def\1{{\sst{(1)}}}
\def\2{{\sst{(2)}}}
\def\3{{\sst{(3)}}}
\def\4{{\sst{(4)}}}
\def\5{{\sst{(5)}}}
\def\6{{\sst{(6)}}}
\def\7{{\sst{(7)}}}
\def\8{{\sst{(8)}}}
\def\m{{\sst{(m)}}}
\def\n{{\sst{(n)}}}
\def\cA{{{\cal A}}}
\def\cB{{{\cal B}}}
\def\cF{{{\cal F}}}
\def\cG{{{\cal G}}}
\def\cH{{{\cal H}}}
\def\tV{\widetilde V}
\def\tW{\widetilde W}
\def\tH{\widetilde H}
\def\tE{\widetilde E}
\def\tF{\widetilde F}
\def\tA{\widetilde A}
\def\im{{{\rm i}}}
\def\tY{{{\wtd Y}}}
\def\ep{{\epsilon}}
\def\vep{{\varepsilon}}
\def\bD{{{\bar D}}}
\def\R{{{\mathbb R}}}
\def\C{{{\mathbb C}}}
\def\H{{{\mathbb H}}}
\def\CP{{{\mathbb C}{\mathbb P}}}
\def\RP{{{\mathbb R}{\mathbb P}}}
\def\Z{{{\mathbb Z}}}
\def\bA{{{\mathbb A}}}
\def\bB{{{\mathbb B}}}
\def\bC{{{\mathbb C}}}
\def\bD{{{\mathbb D}}}
\def\bE{{{\mathbb E}}}
\def\bZ{{{\mathbb Z}}}
\def\Re{{{\frak{Re}}}}
\def\Im{{{\frak{Im}}}}
\def\cosec{{\,\hbox{cosec}\,}}
\def\Gm{{\Gamma_{\!\! -}}}
\def\Gp{{\Gamma_{\!\! +}}}
\def\stan{{standard }}
\def\nonstan{{supernumerary }}
\def\p{{\partial}}
\def\kdel#1{{\fft{\del}{\del#1}}}

\def\bog{{Bogomolny }}
\def\om{{\omega}}

\newcommand{\tamphys}{\it George and Cynthia Woods Mitchell  Institute
for Fundamental Physics and Astronomy,\\
Texas A\&M University, College Station, TX 77843, USA}

\newcommand{\auth}{
Zhao-Long Wang\hoch{\ddagger},
Jianwei Mei\hoch{\dagger} and H. L\"u\hoch{\dagger\star}}

\begin{document}

\begin{flushright}
\hfill{
MIFP-09-01\ \ \ \ USTC-ICTS-09-02}
\end{flushright}

\vspace{25pt}

\begin{center}

{\large {\bf $GL(n,\R)$ Wormholes and Waves in Diverse Dimensions}}

\vspace{25pt}
\auth

\vspace{10pt}

\hoch{\ddagger}{\it Interdisciplinary Center of Theoretical Studies,\\
University of Science and Technology of China, Hefei, Anhui 230026, China}

\vspace{10pt}
\hoch{\dagger}{\tamphys}

\vspace{10pt}

\hoch{\star}{\it Division of Applied Mathematics and Theoretical
Physics,\\
China Institute for Advanced Study,\\
Central University of Finance and Economics, Beijing, 100081, China
}

\vspace{40pt}

\underline{ABSTRACT}
\end{center}

We construct the most general Ricci-flat metrics in $(D+n)$ dimensions
that preserve the $\R^{1,n-1}\times SO(D)$ isometry.  The equations of
motion are governed by the system of a $GL(n,\R)/SO(1,n-1)$ scalar
coset coupled to $D$-dimensional gravity.  Among the solutions, we
find a large class of smooth Lorentzian wormholes that connect two
asymptotic flat spacetimes.  In addition, we obtain new vacuum
tachyonic wave solutions in $D\ge 4$ dimensions, which fit the general
definition of pp-waves in that there exists a covariantly constant
null vector.  The momenta of the tachyon waves are larger than their
ADM masses.  The world-volume of the tachyon wave is $\R^{1,2}$,
instead of $\R^{1,1}$ for the usual vacuum pp-wave.  We show that the
tachyon wave solutions admit no Killing spinors, except in $D=4$, in
which case it preserves half of the supersymmetry.  We also obtain a
general class of $p$-brane wormhole and tachyon wave solutions where
the $\R^{1,n-1}$ part of the spacetime lies in the the world-volume of
the $p$-branes.  These include examples of M-branes and
D3-brane. Furthermore, we obtain AdS tachyon waves in $D\ge 4$
dimensions.

\vspace{15pt}

\thispagestyle{empty}

\pagebreak
\setcounter{page}{1}

\tableofcontents

\addtocontents{toc}{\protect\setcounter{tocdepth}{2}}

\pagebreak

\section{Introduction}

Recently, a large class of smooth Lorentzian wormholes were
constructed in $D\ge 5$ dimensions \cite{lumeiwh}.  These include the
previously-known example in $D=5$ \cite{chde}.  Such a wormhole can be
viewed as a gravitational string that carries a momentum propagating
in one space direction.  The metric has an isometry of $\R^{1,1}\times
SO(D)$ in $(D+2)$ dimensions, where $\R^{1,1}$ denotes the
world-volume, comprising the time and the momentum-carrying space.  If
one performs Kaluza-Klein reduction in the $\R^{1,1}$ directions, the
$D$-dimensional solution is then a spherical symmetric wormhole
supported by a $GL(2,\R)/O(1,1)$ scalar coset \cite{euclid}.

           In this paper, we generalise above construction to obtain
the most general Ricci-flat metrics in $(D+n)$ dimensions that
preserve the $\R^{1,n-1}\times SO(D)$ isometry.  The metric ansatz has
the form
\be
ds_{D+n}^2 = r^2 d\Omega_{D-1}^2 + \fft{dr^2}{f} +
\sum_{\mu,\nu=0}^{n-1} g_{\mu\nu} dz^\mu dz^\nu\,,\label{mostgen}
\ee
where $f$ and $g_{\mu\nu}$ depend only on the $D$-dimensional radial
variable $r$.  Note that this ansatz encompasses also all the
spherical symmetric $p$-branes and pp-waves.  The Kaluza-Klein
reduction on the $z^\mu$'s, which include a time direction, was
performed in \cite{euclid}.  The $D$-dimensional system consists of
the Euclidean-signatured metric and an $GL(n,\R)/SO(1,n-1)$ scalar
coset \cite{stelle,huju, euclid}.  The task is then reduced to
construct spherical symmetric solutions in $D$-dimensions that are
supported by the $GL(n,\R)/SO(1,n-1)$ scalar coset.

     The group $GL(n,\R)$ is a direct product of the $\R$ and
$SL(n,\R)$ groups, where the $\R$ factor is the breathing mode that
measures the overall scale size of the $\R^{1,n-1}$ spacetimes.  It is
consistent to truncate out this mode, leaving the $SL(n,\R)/SO(1,n-1)$
coset.  In sections 2 to 5, we obtain the most general solutions
supported by this scalar coset.  We first present the general
formalism in section 2, then proceed with the simplest $SL(2,\R)$ case
in section 3, and $SL(3,\R)$ in section 4.  The most general solutions
for arbitrary $SL(n,\R)$ were presented in section 5.

        Among the solutions we obtain, there is a large class of new
smooth Lorentzian wormholes.  The role that wormholes play in string
theory has been studied recently in the context of AdS/CFT
correspondence.  In Lorentzian signature, wormholes that connect two
asymptotic AdS spacetimes appear unlikely, and disconnected boundaries
can only be separated by horizons \cite{gsww}.  Thus the recent
studies of wormholes in string theory and in the context of the
AdS/CFT correspondence have so far concentrated on
Euclidean-signatured spaces \cite{mm,bcpvv,hop,bd,bcptv}.  By
combining the standard brane ansatz and wormhole solutions obtained in
\cite{lumeiwh}, a large class of Lorentzian brane wormholes were
obtained in \cite{adswh}.  These solutions include examples of
wormholes that connect AdS$\times$Sphere in one asymptotic region to a
Minkowski spacetime in the other.  In section 7, we obtain analogous
brane solutions for our new Ricci-flat wormholes.

        We also obtain a new vacuum gravitational wave solutions
(\ref{tachyonwave}) and (\ref{tachyonwave45}) for all $D\ge 4$
dimensions.  We verify, up to the cubic order, that the polynomial
scalar invariants of the Riemann tensor vanish identically.  We expect
that, as in the case of the vacuum pp-wave solution, all these
invariants vanish identically.  The solutions fit the general
definition of pp-waves in that there exists a covariantly constant
null vector.  However, the solutions have some distinct features.  One
odd property of our new wave solutions is that the linear momenta are
greater than their AdM masses.  Thus, we call these solutions as
vacuum tachyon waves.  The world-volume of the tachyon wave has three
dimensions, instead of the two dimensions for the pp-wave.  In
appendix B, we show that these tachyon wave metrics admit no Killing
spinors, except for $D=4$, in which case, the solution preserves half
of the supersymmetry.  We also construct $p$-brane solutions with such
a tachyon wave propagating in the world-volume of the brane in section
7.  Furthermore, we obtain AdS tachyon waves in all $D\ge 4$
dimensions.

       In section 6, we construct the most general solution for the
case where the world-volume spacetime $\R^{1,n-1}$ is replaced by
Euclidean space $\R^n$. In section 8, we include the breathing mode
and hence the scalar coset is $GL(n,\R)/SO(1,n-1)$.  We obtain the
most general solutions also for this case.  We conclude our paper in
section 9.  In appendix A, we present the discussion of the properties
of a constant matrix that is crucial for solving the scalar equations
of motion.

\section{General formalism}

    One goal of this paper is to construct the most general Ricci-flat
metrics with an $\R^{1,n-1}\times SO(D)$ isometry in $(D+n)$
dimensions.  The Kaluza-Klein reduction on $\R^{1,n-1}$ gives rise to
a scalar coset of $GL(n,\R)/SO(1, n-1)$ in $D$-dimensions
\cite{euclid}.  In this section, we set up the general formalism for
the case where the breathing mode, associated with the $\R$ factor of
the $GL(n,\R)=\R\times SL(n,\R)$, is (consistently) truncated out.

      We begin by reviewing the $SL(n,\R)/SO(1,n-1)$ scalar coset.  It
can be parameterised by the upper-triangular Borel gauge, which
includes all the positive-root generators $E_i{}^j$ with $i<j$ and
$(n-1)$ Cartan generators $\vec H$.  They satisfy the Borel subalgebra
\be
[\vec H, E_i{}^j] = \vec b_i{}^j E_i{}^j\,,\qquad
[E_i{}^j, E_k{}^\ell] =
\delta^{j}_k E_i{}^\ell - \delta^\ell_i E_k{}^j\,,
\ee
where $\vec b_i{}^j$ are the positive-root vectors.  Following the
general discussion in \cite{cjlp1,cjlp2}, one can parameterise the
$SL(n,\R)/SO(1,n-1)$ coset representative ${\cal V}={\cal V}_1 {\cal
V}_2$, with
\bea
{\cal V}_1 &=& e^{\ft12\vec\phi\cdot \vec H}\,,\nn\\
{\cal V}_2 &=& \prod_{i<j} = \cdots U_{24} U_{23} \cdots
U_{14}U_{13}U_{12}\,,\qquad
U_{ij} \equiv e^{\chi^i{}_j E_i{}^j}\,.
\eea
Here the generators $\vec H$ and $E_i{}^j$ are represented
by $n\times n$ matrices.  Defining a symmetric matrix
\be
{\cal M} = {\cal V}^{\rm T}\, \eta\, {\cal V}\,,\qquad
\eta={\rm diag}(-1,1,\cdots 1)\,,\label{gencalm}
\ee
we propose that the metric ansatz for the $(D+n)$-dimensional
spacetime is given by
\be
ds_{D+n}^2 = ds_D^2 + dz^{\rm T} {\cal M}\, dz\,,\label{genmet1}
\ee
where $dz=(dt,dz_1,\cdots\, dz_{n-1})$.  If we had chosen $\eta$ in
(\ref{gencalm}) to be the identity matrix instead, we would have an
Euclidean-signatured space for the coordinates $t$ and $z_i$'s.  For
the vacuum solution with all the scalars $\vec\phi$ and $\chi^i{}_j$
vanishing, the metric in the $t$ and $z_i$ directions is given by
$ds^2_n=-dt^2 + dz_1^2 + \cdots dz_{n-1}^2$.  We shall refer the
$\R^{1,n-1}$ spacetimes of $t$ and $z_i$ coordinates as the
world-volume of our solutions.  It was shown in \cite{euclid} that the
Kaluza-Klein reductions on the time and on space directions commute.
Thus any permutation of the ``$-1$'' and ``1'' entries in $\eta$ given
by (\ref{gencalm}) is equivalent.

          The Kaluza-Klein reduction on $R^n$ (or $T^n$) with this
type of ansatz, corresponding to $\eta$ being the identity matrix, was
considered \cite{cjlp1,cjlp2}, which is effectively implementing
successively \cite{lpss,lpsoliton,lps} the $S^1$ reduction.  The
reduction on $\R^{1,n-1}$ is analogous, and was performed
in\cite{euclid}.  The $D$-dimensional effective Lagrangian is given by
\be
{\cal L} = \sqrt{g} \Big(R + \ft14 \tr (\del_\mu {\cal M}^{-1}
\del^\mu {\cal M})\Big)\,.\label{genlag}
\ee
Owing to the consistency of the reduction,
the Ricci-flatness of the $(D+n)$-dimensional metric (\ref{genmet1})
becomes equivalent to solutions to the lower $D$-dimensional
system (\ref{genlag}).

     The Lagrangian (\ref{genlag}) is invariant under the global
$SL(n,\R)$ symmetry, with the transformation rule
\be
{\cal M}\longrightarrow  \Lambda^T\,{\cal M}\,
\Lambda\,,\label{slnrtrans}
\ee
where $\Lambda$ is any $n\times n$ constant matrix satisfying
$\det\Lambda=1$.  This symmetry has an origin as rather simple general
coordinate transformations in the $(D+n)$ dimensions. Indeed, the
metric (\ref{genmet1}) is invariant under (\ref{slnrtrans}) together
with $dz\rightarrow \Lambda^{-1} dz$.

       Throughout the paper, we consider solutions that is spherical
symmetric for the $D$-dimensional metric $ds_D^2$.  Without loss of
generality, We take the metric to have the form
\be
ds_D^2=\fft{dr^2}{f} + r^2 d\Omega_{D-1}^2\,,
\ee
where $f$ and all the scalars depend on the radial variable $r$
only.  Einstein equations in the foliating sphere $S^{D-1}$ directions
imply that
\be
\fft{(D-2)(1-f)}{r^2} - \fft{f'}{2r} =0\,,\label{genfeq}
\ee
where a prime denotes a derivative with respect to $r$.  Thus we have
\be
f=1- \Big(\fft{a}{r}\Big)^{2(D-2)}\,.\label{genfsol}
\ee
The Einstein equation associated with the $R_{rr}$ term implies that
\be
-\fft{(D-1)f'}{2r\,f} + \ft14 \tr ({\cal M}^{-1'} {\cal M}')
=0\,.\label{slnrrr}
\ee
The scalar equations of motion are given by
\be
({\cal M}^{-1} \dot {\cal M})\,\dot{} =0\,,\label{slnrscalareq}
\ee
where a dot denotes a derivative with respect to
$\rho$, defined by
\be d\rho=\fft{dr}{r^{D-1}\sqrt{f}}\,.
\ee
The second-order differential equations (\ref{slnrscalareq})
can be easily integrated to give rise to a set of
first-order equations, given by
\be
{\cal M}^{-1} \dot {\cal M} = {\cal C}\,,
\ee
where ${\cal C}$ is a Lie-algebra valued constant matrix.
Substituting this and (\ref{genfsol}) into (\ref{slnrrr}), we have
\be
{\cal I}\equiv - \ft12\tr ({\cal C}^2) = 2(D-1)(D-2) a^{2(D-2)}\,.
\ee
For the solution to be absent from a naked curvature power-law
singularity at $r=0$, it is necessary to have $a^{2(D-2)}\ge 0$, which
implies that ${\cal I} \ge 0$.  For the case with ${\cal I}<0$, and
hence $a^{2(D-2)}<0$, a naked curvature power-law singularity at $r=0$
is unavoidable.  Note that the quantity ${\cal I}$ is invariant under
the global symmetry transformation, under which, ${\cal C}$ transforms
as
\be
{\cal C} \rightarrow \Lambda^{-1} {\cal C} \Lambda\,.
\ee

    It is worth mentioning that the solution for $f$ and equations for
${\cal M}$ apply to any scalar coset ${\cal M}$, not just the
$SL(n,\R)/SO(1,n-1)$ case which we consider.  In the next three
sections, we shall obtain the most general solutions for the
$SL(2,\R)$, $SL(3,\R)$ and $SL(n,\R)$ respectively.

\section{General $SL(2,\R)$ solutions}

  In this section, we give a detail discussion for the case with
$n=2$.  The Borel subalgebra of $SL(2,\R)$ is generated by $H$ and
$E_+$, given by
\be
H=\begin{pmatrix} 1 & 0 \cr 0 & -1\end{pmatrix}\,,\qquad
E_+=\begin{pmatrix} 0 & 1 \cr 0 & 0\end{pmatrix}\,.
\ee
The coset can be parameterised by
\be
{\cal V} = e^{\ft12 \phi H} e^{\chi E_+} =
\begin{pmatrix} e^{\ft12\phi} & \chi e^{\ft12\phi} \cr
0 & e^{-\ft12\phi}\end{pmatrix}\,.
\ee
It follows that ${\cal M}$ is a symmetric $2\times 2$ matrix,
given by
\be
{\cal M}={\cal V}^{\rm T} \eta {\cal V} =\begin{pmatrix}
-e^{\phi} & -\chi e^{\phi}\cr
-\chi e^{\phi} & e^{-\phi} -\chi^2 e^{\phi}
\end{pmatrix}\,,\qquad
\eta ={\rm diag}(-1,1)\,.
\ee
Substitute this to (\ref{genlag}), we have
\be
{\cal L} = \sqrt{g} (R - \ft12 (\del\phi)^2 +
\ft12 e^{2\phi} (\del\chi)^2)\,.
\ee

The equations of motion (\ref{slnrscalareq}) for the scalars
can be solved by the following general ansatz
\be
{\cal M}^{-1} \dot {\cal M} = {\cal C} \equiv\begin{pmatrix}
c_{11} & c_{12}\cr
c_{21} & c_{22}
\end{pmatrix}\label{sl2rfo}
\ee
where $c_{ij}$ are constants subject to the traceless condition
$c_{22}=-c_{11}$.  The equations of (\ref{sl2rfo}) can be written
explicitly, given by
\be
\dot\phi = c_{11} + c_{21} \chi\,,\qquad
\dot \chi = - c_{21} e^{-2\phi}\,,\label{sl2rfo1}
\ee
together with an algebraic constraint
\be
c_{21} e^{-2\phi} - c_{21} \chi^2 - 2 c_{11} \chi + c_{12}=0\,.
\label{sl2r0con}
\ee
It is easy to verify that this algebraic constraint is consistent with
the first-order equations (\ref{sl2rfo1}), and hence a partial
solution to these equations.  Substituting (\ref{sl2rfo1}),
(\ref{sl2r0con}) and (\ref{genfsol}) into (\ref{slnrrr}), we have the
following constraint
\be
{\cal I} =-c_{11}^2 - c_{12} c_{21}=2(D-1)(D-2) a^{2(D-2)}\,.
\label{sl2rreal}
\ee
Thus the solution is parameterised by four parameters, namely
$c_{11},c_{12},c_{21}$ and one extra from solving (\ref{sl2rfo1}) and
(\ref{sl2r0con}).

     As we discussed earlier, the system has an $SL(2,\R)$ global
symmetry which can be used to fix three of the four parameters,
leaving one parameter arbitrary.\footnote{If one takes into account
that the metric is Ricci-flat, and hence it remains a solution with
any constant scaling of the metric, this parameter can be viewed as
trivial as well.}  To do this explicitly, we first note that we can
use the Borel subgroup to diagonalise ${\cal V}$ and hence ${\cal M}$,
at one point in spacetime.  We choose this point to be at asymptotic
infinity $r= \infty$, such that the solutions are asymptotic
Minkowskian with the standard diagonal metric.  This can be achieved
by making use of the Borel transformation, $\phi\rightarrow \phi+ c_1$
and $\chi\rightarrow \chi + c_2$, to set $\phi=0$ and $\chi=0$ at
$r=\infty$, corresponding to ${\cal M}_\infty={\rm diag} (-1,1)$.
After imposing this boundary condition, it follows from
(\ref{sl2r0con}) that we have
\be
c_{21}=-c_{12}\,.
\ee
Thus at asymptotic infinity, we have
\be
\dot{\cal M}\Big|_{r\rightarrow \infty} = \widetilde {\cal C}
= -\begin{pmatrix}
c_{11} & c_{12}\cr
c_{12} & c_{11}
\end{pmatrix}\,.
\ee
We now apply the residual $O(1,1)$ symmetry, whose group elements
can be parameterised as
\be
\Lambda =\begin{pmatrix}
c & s\cr
s & c
\end{pmatrix}\,,\label{o11group}
\ee
where $c=\cosh\delta$ (and $s=\sinh\delta$) is the boost parameter.
This symmetry leaves ${\cal M}_{\infty}={\rm diag}(-1,1)$ invariant,
and transforms $C$ and equivalently $\widetilde {\cal C}$ as follows
\be
{\cal C}\rightarrow \Lambda^{-1} {\cal C}\, \Lambda\,,\qquad
\widetilde {\cal C}\rightarrow \Lambda^{\rm T} \widetilde
{\cal C}\,\Lambda\,.\label{sl2ro11}
\ee
Thus, we have
\be
c_{11}\rightarrow  (c^2+s^2) c_{11} + 2 c s c_{12}
\,,\qquad
c_{12}\rightarrow  (c^2+s^2) c_{12} + 2 c s c_{11}\,.
\ee
Depending on the values of $c_{ij}$, three inequivalent classes arise.

\bigskip
\noindent{\bf Class I: $c_{11} < c_{12}$}
\bigskip

In this class, ${\cal C}$ cannot be diagonalised to a real
matrix.  For simplification, we can use the $O(1,1)$ symmetry to
set $c_{11}=0$ instead.   It follows that we have the following
solution
\be
\chi=\tan\Big(\sqrt{\fft{2(D-1)}{D-2}} \arcsin(\fft{a}{r})^{D-2}\Big)\,,
\qquad e^{2\phi}=\fft{1}{1 + \chi^2}\,.
\ee
The corresponding $(D+2)$-dimensional metric is then given by
\be
ds_{D+2}^2 = \fft{dr^2}{1 - (\fft{a}{r})^{2(D-2)} } +
r^2 d\Omega_{D-1}^2 + \fft{1}{\sqrt{1 + \chi^2}}
\Big(-dt^2 + dz_1^2 - 2\chi\, dt\, dz_1\Big)
\,.
\ee
For $a^{2(D-2)}>0$, this is precisely the Ricci-flat
Lorentzian wormhole solution obtained in \cite{lumeiwh}, although
now with a different radial coordinate. For
$a^{2(D-2)}<0$, the solution has a naked curvature power-law
singularity at $r=0$.

\bigskip
\noindent{\bf Class II: $c_{11} > c_{12}$}
\bigskip

     In this class, we can find a boost parameter $\delta$ to set
$c_{12}=0$, so that the matrix ${\cal C}$ is diagonal.  
We send $a^{2(D-2)}\rightarrow -a^{2(D-2)}$ since in this
case we have ${\cal I}\le 0$.  The solution can be
straightforwardly obtained, given by
\bea
ds_{D+2}^2 &=& r^2 d\Omega_{D-1}^2 + \fft{dr^2}{1 +
(\ft{a}{r})^{2(D-2)}} - e^{\lambda \rho} dt^2 +
e^{-\lambda \rho} dz_1^2\,,\nn\\
\rho&=&-\fft{1}{(D-2)a^{D-2}} {\rm arcsinh} \Big(\fft{a}{r}\Big)
^{D-2}\,,\quad \lambda =\sqrt{2(D-1)(D-2)} a^{D-2}\,.
\eea
The solution has a naked curvature power-law singularity at $r=0$.

\bigskip
\noindent{\bf Class III: $c_{11} = c_{12}$}
\bigskip

      We shall parameterise ${\cal C}$ to be
\be
{\cal C}_2 =\alpha \begin{pmatrix}
1 & 1\cr
-1 & -1
\end{pmatrix}\,,\label{sl2rrank0}
\ee
This $2\times2$ matrix is rank $1$ with all eigenvalues
vanishing; it cannot be diagonalised.  The condition (\ref{sl2rreal})
implies that $a=0$, and hence $f=1$.  The scalars $\chi$ and $\phi$
can be easily solved, given by
\be
e^{\phi}=1 - \fft{q}{r^{D-2}}\,,\qquad
\chi=1 - e^{-\phi} \,,\qquad
q=\fft{\alpha}{D-2}\,.
\ee
The corresponding $(D+2)$-dimensional metric describes the
standard pp-wave, given by
\be
ds_{D-2}^2=-du dv + \fft{q}{r^{D-2}} dv^2 + dr^2 +
r^2 d\Omega_{D-1}^2\,.\label{ppwave}
\ee
where $u=t-z_1$ and $v=t+z_1$ are the asymptotic light-cone coordinates.

     Thus we have demonstrated that there are total three classes of
solutions of the $SL(2,\R)$ system. By requiring the absence of naked
curvature power-law singularity, the most general Ricci-flat metric in
$(D+2)$ dimensions with the $\R^{1,1}\times SO(D)$ isometry with fixed
$\R^{1,1}$ volume is either a wormhole or a pp-wave.  As was observed
in \cite{lumeiwh}, the latter can be obtained as a singular infinite
boost of the former.

      It is worth pointing out that the boosted Schwarzschild black hole
is not include in the solution space.  This is because the volume form
for world-volume $\R^{1,1}$ in this case is not a constant.

\section{General $SL(3,\R)$ solutions}

The coset $SL(3,\R)/SO(1,2)$ representative ${\cal V}$ in a Borel
gauge is given by \cite{euclid}
\bea
{\cal V} &=& e^{\fft12 \vec\phi\cdot\vec H}\, e^{\chi_{23}\, E_{23}}\,
 e^{\chi_{13}\, E_{13}}\,  e^{\chi_{12}\, E_{12}}\nn\\
&=& \begin{pmatrix} e^{\fft1{2\sqrt3}\phi_1 -\fft12\phi_2} &
    \chi_{12}\, e^{\fft1{2\sqrt3}\phi_1 -\fft12\phi_2} &
 \chi_{13}\, e^{\fft1{2\sqrt3}\phi_1 -\fft12\phi_2} \cr
0 & e^{-\fft1{\sqrt3}\phi_1} & \chi_{23}\, e^{-\fft1{\sqrt3}\phi_1} \cr
0 & 0 & e^{\fft1{2\sqrt3}\phi_1 +\fft12\phi_2}
\end{pmatrix} \ ,\label{sl3v}
\eea
where $\vec H$ represents the two Cartan generators, and $E_{ij}$
denote the positive-root generators of $SL(3,\R)$.  It follows that we
have
\be
{\cal M} = {\cal V}^T \eta\, {\cal V}\,,\qquad\qquad
\eta={\rm diag}\, (-1,1,1)\ .\label{sl3calm}
\ee
The $D$-dimensional Lagrangian (\ref{genlag}) is then given by
\bea
(\sqrt{g})^{-1} {\cal L}&=& R - \ft12 (\del\phi_1)^2 - \ft12
(\del\phi_2)^2 + \fft12 e^{-2\phi_2} (\del \chi_{13} -
\chi_{23} \del\chi_{12})^2\nn\\
&&\qquad - \ft12 e^{-\sqrt3 \phi_1 - \phi_2} (\del \chi_{23})^2
+ \ft12 e^{\sqrt3\phi_1 - \phi_2} (\del\chi_{12})^2\,.
\eea
The scalar equations of motion (\ref{slnrscalareq}) can be
solved by the following constant matrix
\be
{\cal M}^{-1} \dot {\cal M} = {\cal C} \equiv\begin{pmatrix}
c_{11} & c_{12} & c_{13}\cr
c_{21} & c_{22} & c_{23}\cr
c_{31} & c_{32} & -c_{11}-c_{22}
\end{pmatrix}\,.\label{sl3rfo}
\ee
One can read off five first-order equations for the five scalars
$\phi_1$, $\phi_2$, $\chi_{12}$, $\chi_{13}$ and $\chi_{23}$.  In
addition, there are three algebraic constraints which are partial
solutions to the equations.  Substituting (\ref{sl3rfo}) to
(\ref{slnrrr}), we obtain
\bea
{\cal I} &=& -\ft12 \tr({\cal C}^2) =
-(c_{11}^2 + c_{22}^2 + c_{12} c_{21} + c_{11} c_{22} +
c_{13}c_{31} + c_{23}c_{32})\nn\\
&=& 2(D-1)(D-2) a^{2(D-2)}\,.
\label{sl3rcon}
\eea
Thus we see that the solution is parameterised by a total of 10
constant parameters, eight $c_{ij}$'s and two extra from solving for
the first-order equations and algebraic constraints.  The system has
an $SL(3,\R)$ global symmetry, which can remove eight parameters,
leaving the solution with two arbitrary parameters.

       To apply the global $SL(3,\R)$ symmetry, we first use the Borel
transformations to set $\phi_i=0$ and $\chi_{ij}=0$ at the asymptotic
infinity, $r=\infty$.  The spacetimes at $r=\infty$ is then
Minkowskian since we have ${\cal M}_\infty=\eta$.  Apply this boundary
condition to the three algebraic constraints, we find that
\be
c_{21}=-c_{12}\,,\qquad
c_{31}=-c_{13}\,,\qquad
c_{32}=c_{23}\,.
\ee
Thus we have
\be
\dot {\cal M}\Big|_{r\rightarrow \infty}\equiv \widetilde {\cal C}=
- \begin{pmatrix}
c_{11} & c_{12} & c_{13}\cr
c_{12} & -c_{22} & - c_{23}\cr
c_{13} & -c_{23} & c_{11}+ c_{22}
\end{pmatrix}\,,
\ee
The upshot of this discussion is that the constant traceless matrix
${\cal C}$ defined in (\ref{sl3rfo}), after applying the Borel
subgroup of $SL(3,\R)$ global transformations, can be expressed as
${\cal C}=\eta \widetilde {\cal C}$, where $\widetilde {\cal C}$ a
symmetric constant matrix.

     As we shall discuss in detail in appendix for the general
$SL(n,\R)$, three different classes of solutions emerge depending on
the value of the matrix ${\cal C}$.  In this section, we shall just
present the results.

\bigskip
\noindent{\bf Class I:}
\bigskip

The first class corresponds to the case where
${\cal C}$ has a pair of complex eigenvalues.  It is isomorphic
to
\be
{\cal C} = \begin{pmatrix}
\alpha & -\beta & 0\cr
\beta & \alpha & 0\cr
0 & 0 & -2\alpha
\end{pmatrix}\,.
\ee
We find that the corresponding $(D+3)$-dimensional metric is given by
\bea
ds^2 &=& r^2 d\Omega_{D-1}^2 +
\fft{dr^2}{1 - (\ft{a}{r})^{2(D-2)}}\nn\\
&& e^{\alpha \rho} [\cos(\beta\rho) (-dt^2 + dz_1^2) +
2\sin(\beta\rho)\, dt\, dz_1] + e^{-2\alpha \rho} dz_2^2\,,\nn\\
\rho&=& -\fft{1}{(D-2)a^{D-2}} \arcsin\Big(\fft{a}{r}\Big)
^{D-2}\,,
\eea
with
\be
\beta^2 - 3\alpha^2 = 2(D-1)(D-2) a^{2(D-2)}\,.
\ee
For $a^{2(D-2)}>0$, the solution describes a
smooth Lorentzian wormhole.  In special case with $\alpha=0$, it
describes a direct product of an $SL(2,\R)$ wormhole, discussed in
section 3, with a real line. For $a^{2(D-2)}<0$, the
solution has a naked curvature power-law singularity at $r=0$.
When $\beta^2=3\alpha^2$, corresponding to have
$a=0$, the solution has a naked curvature power-law singularity at
$r=0$, since in this case $\rho\sim 1/r^{D-2}$.

\bigskip
\noindent{\bf Class II:}
\bigskip

The second class corresponds to that ${\cal C}$ has three real
eigenvalues, with one time-like and 2 space-like eigenvectors.  The
matrix ${\cal C}$ can be diagonalised by a certain $SO(1,2)$ matrix to
give ${\cal C} = {\rm diag}(\lambda_0, \lambda_1, \lambda_2)$ with
$\lambda_2=-\lambda_0 -\lambda_1$. In this case, we have ${\cal I} =
-\ft12 \lambda_0^2 - \ft12\lambda_1^2 -\ft12 (\lambda_0 +
\lambda_1)^2$.  It follows from (\ref{sl3rcon}) that the reality
condition implies that $f=1 + a^{2(D-2)}/r^{2(D-2)}$.  Thus we find
that the corresponding $(D+3)$-dimensional metric is given by
\bea
ds^2 &=& r^2 d\Omega_{D-1}^2 + \fft{dr^2}{1 + (\ft{a}{r})^{2(D-2)}}
- e^{\lambda_0 \rho } dz_0^2 + e^{\lambda_1 \rho} dz_1^2
+ e^{-(\lambda_0+\lambda_1)\rho} dz_3^2\,,\nn\\
\rho &=& -\fft{1}{(D-2)a^{D-2}} {\rm arcsinh}\Big(\fft{a}{r}\Big)^{D-2}\,,
\eea
where
\be
\lambda_0^2 + \lambda_1^2 + (\lambda_0 +\lambda_1)^2 =
4(D-1)(D-2) a^{2(D-2)}\,.
\ee
This solution has a naked curvature power-law singularity at $r=0$.

\bigskip
\noindent{\bf Class III:}
\bigskip

      In the third class, the matrix ${\cal C}$ has three real
eigenvalues, but with degenerate eigenvectors.  There are two
inequivalent cases.  The first case corresponds to that ${\cal C}$ is
of rank 2 and all of its eigenvalues vanish. Such a ${\cal C}$ is
isomorphic to
\be
{\cal C}_3 = \alpha \begin{pmatrix}
\cos\beta & -\cos\beta & \ft12 \sin\beta \\
-\cos\beta & -\cos\beta &   \ft12\sin\beta\\
-\ft12\sin\beta & \ft12\sin\beta & 0\\
\end{pmatrix}\,.\label{sl3rrank0}
\ee
We find that the corresponding $(D+3)$-dimensional metric is given by
\bea
ds_{D+3}^2 &=&r^2d\Omega_{D-1}^2+dr^2+d\tilde s_3^2\cr
d\tilde s_3^2&=& - (1 + \rho\cos\beta - \ft18 \rho^2\sin^2\beta)
dt^2 + (1 - \rho \cos\beta + \ft18 \rho^2 \sin^2\beta) dz_1^2\nn\\
&&+2(\rho\cos\beta - \ft18\rho^2 \sin^2\beta) dt dz_1 -
\rho \sin\beta\, dt dz_2\nn\\
&& +\rho\sin\beta\, dz_1 dz_2 + dz_2^2\,.
\eea
with $\rho=-\ft{\alpha}{(D-2)r^{D-2}}$. In the asymptotic light-cone
coordinates $u=t+z_1$ and $v=t-z_1$, the metric becomes
\bea
ds_{D+3}^2 = -du dv + dw^2
-\left(\rho \cos\beta - \ft18 \rho^2 \sin^2\beta\right) dv^2 -
\rho \sin\beta\, dv dw + dr^2 + r^2 d\Omega_{D-1}^2
\,.\label{tachyonwave}
\eea
Here we rename the $z_2$ coordinate to be $w$.  This metric has
non-vanishing Riemann tensor components, and hence it is not flat.  We
also verify that the Riemann tensor square and cubic scalar invariants
all vanish identically.  We expect that as in the case of the vacuum
pp-wave solution, all the polynomial scalar invariants of the Riemann
tensor for the metric (\ref{tachyonwave}) vanish identically.

     The new wave solutions fit the general definition of pp-waves in
that there exists a covariantly constant null vector $k=\del/\del_u$.
However, there are several differences comparing this new wave
solution to the usual pp-wave.  The world-volume for the usual pp-wave
solution is two dimensional, whilst it has three dimensions spanned by
$(t,z_1, z_2)$ (or $(u,v,w)$) for the new solution.  The mass and the
momentum are given by
\be
M=\alpha \cos\beta\,,\qquad
P_1 = \alpha\cos\beta\,,\qquad
P_2 = -\ft12 \alpha \sin\beta\,,\label{massp}
\ee
Note that the mass and the momentum components are evaluated from the
Komar integrals for the Killing vectors $\del/\del t$ and $\del/\del
z_i$ respectively.  We omitted certain overall inessential constant
factor in presenting the quantities above.  Thus the solution is of
tachyonic nature, since $M^2-P_1^2-P_2^2=- \ft14 \alpha^2 \sin^2\beta
\le 0$.  We shall call this solution tachyon wave.  We can make an
orthonormal transformation
\be
z_1\rightarrow \fft{2\cos\beta\, z_1 + \sin\beta z_2}{
\sqrt{4\cos^2\beta + \sin^2\beta}}\,,\qquad
z_2\rightarrow \fft{2\cos\beta\, z_2 - \sin\beta z_1}{
\sqrt{4\cos^2\beta + \sin^2\beta}}\,,
\ee
such that the momentum has only the $z_1$ component. In these
coordinates, we have
\be
M=\alpha \cos\beta\,,\qquad
P_1 = \alpha\cos\beta \sqrt{1 + \ft14 \tan^2\beta}\,,\qquad
P_2 =0\,,\label{massp1}
\ee

    The vacuum pp-wave (\ref{ppwave}) has half of Killing spinors.  As
we show in the appendix B, there is no Killing spinor in this tachyon
wave solution.  This is consistent with the fact that the BPS
condition for a tachyon is obviously not satisfied.  When $\beta=0$,
we have $P_2=0$ and $M=P_1$, the BPS condition is then satisfied and
the solution becomes the pp-wave.  When $\beta=\ft12\pi$, the solution
becomes massless, but with non-vanishing linear momentum.

     The tachyon wave metric (\ref{tachyonwave}) is valid also for
$D=1$ and $D=2$, corresponding to total 4 and 5 spacetime dimensions.
The corresponding $\rho$ is given by $\rho=\alpha \log(r)$ and
$\rho=\alpha r$ respectively.  The metrics for these two cases are
given by
\bea
ds_4^2 &=& -du dv + dw^2
-(x\,\alpha \cos\beta- \ft18 x^2 \alpha^2 \sin^2\beta) dv^2 -
x\, \alpha  \sin\beta\, dv dw + dx^2\,,\cr
ds_5^2 &=& -du dv + dw^2
-(\log r\,\alpha \cos\beta- \ft18 (\log r)^2\alpha^2 \sin^2\beta)
dv^2 \cr
&&\qquad -\log r\,\alpha  \sin\beta\,dv dw + dr^2 + r^2 d\phi^2
\,,\label{tachyonwave45}
\eea
However, in these two cases, there is no well-defined asymptotic
region.  Note that the four-dimensional metric belongs to the type N
in the Petrov classification. When $\beta=0$, the four-dimensional
metric is flat.

        We can perform Kaluza-Klein reduction on the $w$ direction
for the tachyon wave (\ref{tachyonwave}). We have
\bea
ds_{D+2}^2 &=& -du dv  - \left(\rho \cos\beta + \ft18 \rho^2
\sin^2\beta\right) dv^2 + dr^2 + r^2 d\Omega_{D-1}^2\,,\cr
{\cal A} &=& -\ft12 \rho \sin\beta\, dv\,.
\eea
Thus, the lower-dimensional solution is a usual pp-wave supported by
a Maxwell field.

      The second degenerate case for ${\cal C}$ is given by
\be
{\cal C} = \begin{pmatrix}
\alpha+\beta & \alpha & 0\cr
-\alpha & -\alpha+\beta & 0 \cr
0 & 0 &-2\beta
\end{pmatrix}\,.\label{c3case2}
\ee
The solution is given by
\bea
ds_{D+3}^2\!\!\!&=&\!\!\!r^2d\Omega_{D-1}^2+
\fft{dr}{1+\left(\ft{a}{r}\right)^{2(D-2)}}
+e^{\beta\rho}(-du dv + q\rho dv^2) + e^{-2\beta\rho}dz_2^2\,,\cr
\rho\!\!\!&=&\!\!\!-\fft1{(D-2)a^{D-2}} {\rm arcsinh}
\left(\frac{a}{r}\right)^{(D-2)}\,,\quad
\beta=\sqrt{\ft23 (D-1)(D-2)}\, a^{2(D-2)} .
\eea
When $\beta=0$, the matrix (\ref{c3case2}) reduces to
(\ref{sl2rrank0}) and the metric above becomes the vacuum pp-wave.
When $\alpha=0$, corresponding to $q=0$, the solution becomes that of
class I.  In general the solution has a naked curvature power-law
singularity at $r=0$.

\section{General $SL(n,\R)$ solutions}

Having discussed the $SL(2,\R)$ and $SL(3,\R)$ examples, it is
straightforward to generalise to obtain the most general spherical
symmetric solutions for any $SL(n,\R)$.  We can use the Borel
transformation to to diagonalise ${\cal M}$ at the asymptotic infinity
$r=\infty$, namely ${\cal M}_\infty=\eta$.  This implies that the
traceless matrix $\cal C$ has the property that $\eta {\cal C}$ is
symmetric.  To be specific, for ${\cal M}$ give by (\ref{gencalm}),
${\cal C}$ is given by
\be {\cal C}=\begin{pmatrix} -\sum_{i=1}^{n-1} \dot
\Phi_i & -\dot \chi_{01} & \dots & -\dot \chi_{0,n-1}\cr
\dot \chi_{01}& \dot \Phi_1 & \dots  & \dot \chi_{1,n-1}\\
\vdots & \vdots & \ddots & \vdots\cr
\dot \chi_{0,n-1} & \dot \chi_{1,n-1} &
\dots &
\dot \Phi_{n-1}\end{pmatrix}|_{r\rightarrow \infty}\,.\label{gencmatrix}
\ee

    As we demonstrate in appendix A, three classes of solution emerge
depending on the values of the ${\cal C}$.

\bigskip
\noindent{\bf Class I:}
\bigskip

The first class corresponds to a ${\cal C}$ with a pair of
complex eigenvalues.  Using the $SO(1,1-n)$ residual global
symmetry, we can simplify ${\cal C}$ as follows
\be%
{\cal C}=\begin{pmatrix}
\alpha & -\beta \cr%
\beta   & \alpha \cr%
& & \lambda_2 \cr%
& & & \ddots \cr%
& & & & \lambda_{n-1}
\end{pmatrix}~,
\ee%
with $\tr({\cal C})=0$.  We find that the corresponding
$(D+n)$-dimensional Ricci-flat metric can be straightforwardly
obtained, given by
\bea
ds_{D+n}^2 &=& r^2 d\Omega_{D-1}^2 + \fft{dr^2}{1 -
\left(\ft{a}{r}\right)^{2(D-2)}}\nn\\
&& + e^{\alpha\rho} [\cos(\beta\rho) (-dt^2 +dz_1^2) +
2\sin(\beta\rho) dt dz_1] + \sum_{i=2}^{n-1}
\lambda^{\lambda_i \rho} dz_i^2\,,\cr
\sum_{i=2}^{n-1} \lambda_i\!\!\! &+&\!\!\! 2\alpha = 0\,,\qquad
2\beta^2-2\alpha^2 -\sum_{i=2}^{n-1} \lambda_i^2 =
4(D-1)(D-2) a^{2(D-2)}\,,\cr
\rho &=&-\fft{1}{(D-2) a^{D-2}} \arcsin\left(\fft{a}{r}\right)
^{D-2}\,.
\eea
For $a^{2(D-2)}>0$, the solution describes a smooth Lorentzian
wormhole. For $a^{2(D-2)}\le 0$, the solution has a naked curvature
power-law singularity at $r=0$.  The global structure and the
traversability of the $SL(2,\R)$ Lorentzian wormhole in $D=5$ were
discussed in detail in \cite{aagc,aagccf,lumeiwh}.  Note that there
are no further off-diagonal terms in $SL(n,\R)$ wormholes than the one
with $SL(2,\R)$.  We expect that the property of the general wormholes
is analogous to the $SL(2,\R)$ one.  We shall discuss their properties
in detail in a future publication.

\bigskip
\noindent{\bf Class II:}
\bigskip

The second class corresponds to a ${\cal C}$ with all real eigenvalues
and one of the eigenvectors being time-like.  In this case, it can be
diagonalised by an $SO(1,n-1)$ transformation.  We thus have
\be
{\cal C}={\rm diag}(\lambda_0, \lambda_1,\cdots, \lambda_{n-1})\,,
\ee
with $\tr({\cal C})=0$.  We find that the corresponding
$(D+n)$-dimensional Ricci-flat metric solution is
\bea
ds_{D+n}^2 &=& r^2 d\Omega_{D-1}^2 + \fft{dr^2}{1 +
\left(\ft{a}{r}\right)^{2(D-2)}} - e^{\lambda_0 \rho} dt^2 +
\sum_{i=1}^{n-1} e^{\lambda_i\rho} dz_i^2\,,\nn\\
\sum_{i=0}^{n-1}\lambda_i&=&0\,,\qquad
\sum_{i=0}^{n-1}\lambda_i^2 = 4(D-1)(D-2) a^{2(D-2)}\,,\nn\\
\rho &=& -\fft{1}{(D-2)a^{D-2}} {\rm arcsinh} \left(
\fft{a}{r}\right)^{D-2}\,.
\eea
The solution is asymptotic Minkowskian; it has a naked
curvature power-law singularity at $r=0$.

\bigskip
\noindent{\bf Class III:}
\bigskip

In this class, All the eigenvalues of the matrix ${\cal C}$ are real,
but there are no time-like eigenvector.  As we demonstrate in Appendix
A, there are two cases of ${\cal C}$.  They can be constructed from
the rank-1 ${\cal C}_2$ and rank-2 ${\cal C}_3$ given in
(\ref{sl2rrank0}) and (\ref{sl3rrank0}) respectively.

       The first case is given by
\be
{\cal C} = \begin{pmatrix}
{\cal C}_2 + \lambda_0 \oneone &  &  &&  \cr
 &\lambda_2 &  &&  \cr
 &  & \lambda_3 &&  \cr
& & &\ddots & \cr
&&&& \lambda_{n-1}
\end{pmatrix}\,,
\ee
with $\tr({\cal C})=0$.  We find that the corresponding
$(D+n)$-dimensional Ricci-flat metric is given by
\bea
ds_{D+n}^2&=&r^2d\Omega_{D-1}^2+
\fft{dr}{1+\left(\ft{a}{r}\right)^{2(D-2)}}
+e^{\lambda_0\rho}(-du dv + q\rho dv^2) + \sum_{i=2}^{n-1}
e^{\lambda_i \rho}dz_i^2\,,\cr
\rho&=&-\fft1{(D-2)a^{D-2}} {\rm arcsinh}
\left(\frac{a}{r}\right)^{(D-2)}\,,\cr
2\lambda_0\!\!\! &+&\!\!\! \sum_{i=2}^{n-1} \lambda_i=0\,,\qquad
2\lambda_0^2 +\sum_{i=2}^{n-1} \lambda_i^2 = 4(D-1)(D-2) a^{2(D-2)}\,.
\eea
The solution reduces to the standard pp-wave when all $\lambda_i$'s
vanish.  In general the solution has a naked curvature
power-law singularity at $r=0$.

    In the second case, we have
\be
{\cal C} = \begin{pmatrix}
{\cal C}_3 + \lambda_0 \oneone &&&& \cr
&\lambda_3 &&&\cr
&& \lambda_4 & & \cr
&&&\ddots &\cr
&&&& \lambda_{n-1}
\end{pmatrix}\,,
\ee
again with $\tr({\cal C})=0$.  We find that the corresponding
$(D+n)$-dimensional Ricci-flat metric is given by
\bea
ds_{D+n}^2 &=& r^2 d\Omega_{D-1}^2+ \fft{dr^2}{1 + \left(\fft{a}{r}
\right)^{D-2}}\nn\\
&&\!\!\!+e^{\lambda_0\rho} [-du dv + dz_2^2
-(\rho \cos\beta - \ft18 \rho^2 \sin^2\beta) dv^2 -
\rho \sin\beta\, dv dz_2] + \sum_{i=3}^{n-1} e^{\lambda_i \rho}dz_i^2
\,,\cr
\rho&=&-\fft1{(D-2)a^{D-2}} {\rm arcsinh}
\left(\frac{a}{r}\right)^{(D-2)}\,,\cr
3\lambda_0\!\!\! &+&\!\!\! \sum_{i=3}^{n-1} \lambda_i=0\,,\qquad
3\lambda_0^2 +\sum_{i=3}^{n-1} \lambda_i^2 = 4(D-1)(D-2) a^{2(D-2)}\,.
\eea
In general the solution also has a naked curvature
power-law singularity at $r=0$.  The tachyon wave arises when
all $\lambda_i$'s vanish.

To summarise, in this section, we obtain the complete set of
spherical symmetric solutions for the $SL(n,\R)/SO(1,n-1)$ coset
scalar, and we obtain the corresponding $(D+n)$-dimensional
Ricci-flat metrics.

\section{Euclidean signature solutions}

In the previous sections, we consider solutions that are asymptotic
Minkowskian.  The solution space becomes much simpler if we choose the
Euclidean signature. In this case, the matrix ${\cal C}$ is symmetric
and can always be diagonalised by an $SO(n)$ transformation.  The most
general solution is then given by
\bea
ds_{D+3}^2&=&r^2d\Omega_{D-1}^2+\fft{dr^2}{1+\left(\frac{a}{r}\right)
^{2(D-2)}}+\sum^{n}_{i=1} e^{\a_i\rho}dz_i^2
\eea
with
\bea
\sum^{n}_{i=1}\a_i&=&0\,,\qquad \sum^{n}_{i=1}
\a_i^2=4(D-1)(D-2)a^{2(D-2)}\,,\cr
\rho&=&-\fft{1}{(D-2)a^{D-2}} {\rm arcsinh}
\left(\frac{a}{r}\right)^{(D-2)}
\eea

\section{General $p$-brane wormholes and waves}

       In sections 2--5, we constructed the most general Ricci-flat
metrics in $(D+n)$ dimensions with the $\R^{1,n-1}\times SO(D)$ isometry.
There are three different classes of such solutions.  Here we
construct charged solutions to $\hat D$-dimensional Einstein gravity
coupled to a $(p+2)$-form field strength, together with a dilaton.
The Lagrangian has the following general form
\be
{\cal L}_{\hat D}=\sqrt{-g} \Big(R- \ft12 (\del\phi)^2 -
\fft{1}{2\,(p+2)!} e^{\alpha\phi} F_{p+2}^2\Big)\,.\label{genbranelag}
\ee
where $F_{p+2}=dA_{p+1}$.  The constant $\alpha$ can be parameterised
as \cite{lpss}
\be
\alpha^2= \Delta - \fft{2(p+1)(\hat D-p-3)}{\hat D-2}\,.\label{avalue}
\ee
The Lagrangian (\ref{genlag}) is of the form that typically arises
as a truncation of the full Lagrangian in many supergravities,
with $\Delta$ being given by
\be
\Delta=\fft{4}{N}\,,
\ee
for integer $N$.  The values of $N$ that can arise depends on
the spacetime dimensions; they are classified in \cite{lpsoliton}.

    We start with the Ricci-flat solutions in $(D+n)$ dimensions
constructed in the paper, and augment the spacetime with flat
directions to become $\hat D=D+n+m$ dimensions.  The Ricci-flat $\hat
D$-dimensional metric has the form
\be
ds_{\hat D}^2 = ds_D^2 + ds_n^2 + \sum_{i=1}^m (dx^i)^2\,,
\ee
where the first two terms denotes the general $SL(n,\R)$ solutions we
obtained in the previous sections.  We now follow \cite{adswh}, and
consider an electric $p$-brane solution where $p=m+n-1$.  We find that
the solution is given by
\bea
ds_{\hat D}^2 &=& H^{\ft{(p+1)N}{\hat D-2}} ds_D^2 +
H^{-\ft{(D-2)N}{\hat D-2}} \Big(ds_n^2 + \sum_{i=1}^m (dx^i)^2
\Big)\cr
e^{\alpha \phi} &=& H^{2 - \fft{(p+1)(D-2)N}{\hat D-2}}\,,\cr
F_{p+2} &=& \sqrt{N} dz_1\wedge\cdots \wedge dz_n\wedge dx_1
\wedge \cdots \wedge dx_m \wedge dH^{-1}\,,\label{pbrane}
\eea
where $H$ is a harmonic function on the metric $ds_D^2$.  Thus
for spherical symmetric branes, we have
\bea
f=1:&& H = 1 + \fft{Q}{r^{D-2}}\,,\cr
f=1-\left(\fft{a}{r}\right)^{2(D-2)}: &&
H = 1 + Q\,\arcsin \left(\fft{a}{r}\right)\,,\cr
f=1+\left(\fft{a}{r}\right)^{2(D-2)}: &&
H = 1 + Q\, {\rm arcsinh} \left(\fft{a}{r}\right)\,.
\eea

   Note that we can also consider ``magnetic $p$-brane wormholes,''
which are equivalent to the previously-discussed electric cases, but
constructed using the $(\hat D-p-2)$-form dual of the $(p+2)$-form
field strength $F_{p+2}$. In other words, we can introduce the dual
field strength
\be
\wtd F_{\td p+2} = e^{\alpha\phi}\, {*F_{p+2}}\,,
\ee
where $\td p=\hat D-p-4$,
in terms of which the Lagrangian (\ref{genbranelag}) can be rewritten as
\be
{\cal L}_{\hat D}=\sqrt{-g} \Big(R- \ft12 (\del\phi)^2 -
\fft{1}{2\,(\td p+2)!} e^{-\alpha\phi} \wtd F_{\td p+2}^2\Big)
\,.\label{genlagdual}
\ee
where $\wtd F_{\td p+2}=d\wtd A_{\td p+1}$.  The electric solution
(\ref{pbrane}) of (\ref{genbranelag}) then be reinterpreted as a
magnetic solution of (\ref{genlagdual}), with $\wtd F_{\sst{(\td n)}}$
given by
\be
\wtd F_{\tilde p+2} = \sqrt{N}\,(D-2)\, Q\, \Omega_{D-1}\,.
\ee

      A particularly interesting class of brane solutions are those
where the dilaton decouples.  These include the M-branes and
D3-brane.  The global structure of such a brane with an $SL(2,\R)$
wormhole was discussed in \cite{adswh}.  These solutions connect
AdS$\times$Sphere in one asymptotic region to a flat Minkowski
spacetimes in the other.  We expect that this property maintains
when the general allowed $SL(n,\R)$ wormholes are added to the branes.

      Here we present explicitly the M-branes and D3-brane on the
background of the tachyon wave (\ref{tachyonwave}) we constructed in
section 4.  We find that the M2-brane tachyon wave is given by
\bea
ds_{11}^2 &=& H^{-\ft23} \Big(-du dv + dw^2 -
(\fft{\alpha\cos\beta}{r^6} -\fft{\alpha^2\sin^2\beta}{8r^{12}})
dv^2 - \fft{\alpha\sin\beta}{r^6} dv dw\Big)\cr
&& +H^{\ft13} (dr^2 + r^2 d\Omega_7^2)\,,\cr
F_\4 &=& du\wedge dv\wedge dw \wedge dH^{-1}\,,\qquad
H=1 + \fft{Q}{r^6}\,.
\eea
The M5-brane tachyon wave is
\bea
ds_{11}^2 &=& H^{-\ft13} \Big(-du dv + dw^2 -
(\fft{\alpha\cos\beta}{r^3} -\fft{\alpha^2\sin^2\beta}{8r^{6}})
dv^2 - \fft{\alpha\sin\beta}{r^3} dv dw+ dx^i dx^i\Big)\cr
&& +H^{\ft23} (dr^2 + r^2 d\Omega_4^2)\,,\cr
F_\4 &=& 3Q\, \Omega_\4\,,\qquad
H=1 + \fft{Q}{r^3}\,.
\eea
Finally, the D3-brane tachyon wave of the type IIB supergravity is
\bea
ds_{10}^2 &=& H^{-\ft12} \Big(-du dv + dw^2 -
(\fft{\alpha\cos\beta}{r^4} -\fft{\alpha^2\sin^2\beta}{8r^{8}}) dv^2
- \fft{\alpha\sin\beta}{r^4} dv dw+ dx^i dx^i\Big)\cr
&& +H^{\ft12} (dr^2 + r^2 d\Omega_5^2)\,,\cr
F_\4 &=& 4Q\, (\Omega_\5 + {*\Omega_\5})\,,\qquad
H=1 + \fft{Q}{r^4}\,.
\eea
In the decoupling limit where the constant 1 in function $H$ can be
dropped, the metric becomes a direct product of AdS tachyon wave
and a sphere.  The AdS tachyon wave in arbitrary dimensions is
given by
\be
ds_{D}^2 = \fft{dy^2}{y^2} + y^2
\Big(-du dv + dw^2 - (\fft{\alpha\cos\beta}{y^{D-1}}
-\fft{\alpha^2\sin^2\beta}{8y^{2(D-1)}}) dv^2 -
\fft{\alpha\sin\beta}{y^{D-1}} dv dw+ dx^i dx^i\Big)\,,
\ee
where $i=1,\ldots,(D-4)$, and the cosmological constant is scaled to
be unit.  In other words, the metrics satisfy $R_{\mu\nu} = - (D-1)
g_{\mu\nu}$.  At asymptotic infinity $r=\infty$, the metric is the
AdS$_D$ in holospherical coordinates.  The solution describes a
tachyon wave propagating in the AdS$_D$ spacetime.  It is easy to
verify that the Polynomial scalar invariants of the Riemann tensors
are all independent on the parameters $\alpha$ and $\beta$.  For
$\beta=0$, it becomes a pp-wave in the AdS.  For $\beta=\ft12\pi$, the
tachyon wave is massless.

\section{Generalise to $GL(n,\R)$}

In the $\R^{1,n-1}$ reduction we considered in section 2, we can also
turn on the breathing mode that scales the volume of the $\R^{1,n-1}$
spacetime.  The resulting scalars in $D$-dimensional parameterise a
coset of $GL(n,\R)/O(1,n-1)$.  The reduction ansatz is given by
\be
ds_{D+n}^2 = e^{-\sqrt{\ft{2n}{(D-2) (D+n-2)}}\, \varphi} ds_D^2 +
e^{\sqrt{\ft{2(D-2)}{n(D+n-2)}}\, \varphi} dz^{\rm T} {\cal M} dz\,,
\ee
The $D$-dimensional Lagrangian is given by
\be
{\cal L} = \sqrt{g}\left(R -\ft12(\del\varphi)^2 +
\ft14 \tr (\del_\mu {\cal M}^{-1} \del^\mu {\cal M})\right)\,.
\ee
We now define
\be
\hat {\cal M} = e^{\ft{1}{\sqrt{2n}}\,\varphi} {\cal M}\,.
\ee
The Lagrangian becomes
\be
{\cal L} = \sqrt{g}\left(R + \ft14 \tr (\del_\mu \hat {\cal M}^{-1}
\del^\mu \hat {\cal M})\right)\,.
\ee
It follows that the constant matrix $\hat {\cal C}$, define by
\be
\hat {\cal M}^{-1} \dot {\hat {\cal M}} \equiv \hat {\cal C}\,,
\ee
has the same property as ${\cal C}$ except that it is no longer
traceless.  The classification of the ${\cal C}$ matrix also
holds for $\hat {\cal C}$.  This leads to the following three
classes of solutions.

\bigskip
\noindent{\bf Class I:}
\bigskip

This generalises the first class of the $SL(n,\R)$
solutions. It is given by 
\bea
ds^2_{D+n} &=& e^{-\ft{\lambda\rho}{D-2}} \Big(
r^2 d\Omega_{D-1}^2 + \fft{dr^2}{1 - \left(\ft{a}{r}\right)^{2(D-2)}}
\Big)\cr
&&+e^{\alpha \rho} [ \cos(\beta\rho) (-dt^2 + dz_1^2) +
2\sin(\beta\rho) dt dz_1] + \sum_{i=2}^{n-1} e^{\lambda_i\rho} dz_i^2
\,,
\eea
where
\bea
&&\rho=-\fft{1}{(D-2)a^{D-2}}\arcsin\left(\fft{a}{r}\right)^{D-2}
\,,\qquad 2\alpha + \sum_{i=2}^{n-1} \lambda_i = \lambda\,,\cr
&&2(\beta^2 - \alpha^2) -\sum_{i=2}^{n-1}\lambda_i^2 -
\fft{\lambda^2}{D-2} = 4(D-1)(D-2) a^{2(D-2)}\,. \eea
For $a^{2(D-2)}>0$, the solution describes a smooth
Lorentzian wormhole that connects two asymptotic flat spacetimes.
For $a^{2(D-2)}\le 0$, the solution has a naked curvature
power-law singularity at $r=0$.

\bigskip
\noindent{\bf Class II:}
\bigskip

This generalises the second class of the $SL(n,\R)$
solutions. The $\hat C$ can be diagonalised, and there is no
traceless condition on the eigenvalues $\lambda_i$. We find that the
$(D+n)$-dimensional Ricci-flat metric is given by
\bea ds_{D+n}^2 &=& e^{-\ft{\lambda\rho}{D-2}} \Big( r^2
d\Omega_{D-1}^2 + \fft{dr^2}{1 + \left(\ft{a}{r}\right)^{2(D-2)}}
\Big) - e^{\lambda_0\rho} dt^2 + \sum_{i=1}^{n-1} e^{\lambda_i \rho}
dz_i^2\,,\nn\\
\sum_{i=0}^{n-1} \lambda_i &=& \lambda\,,\qquad \sum_{i=0}^{n-1}
\lambda_i^2 + \fft{\lambda^2}{D-2} =
4 (D-1)(D-2) a^{2(D-2)}\,,\nn\\
\rho &=& - \fft{1}{(D-2)a^{D-2}} {\rm arcsinh}\left(
\fft{a}{r}\right)^{D-2}\,. \eea
In general, the solution has a naked singularity at $r=0$.  However,
when $\lambda_i=0$ for all $i$ except $\lambda_0$, we have
$\lambda_0=\lambda=2(D-2)a^{D-2}$.  The prefactor of
$d\Omega_{D-1}^2$ becomes a finite and non-vanishing constant at
$r=0$.  The solution is the Schwarzschild black hole with the
horizon located at $r=0$.

\bigskip
\noindent{\bf Class III:}
\bigskip

This generalises the third class of the $SL(n,\R)$ solutions.
The $(D+n)$-dimensional Ricci-flat metrics are given by
\bea
ds_{D+n}^2 &=& e^{-\ft{\lambda\rho}{D-2}} \Big(
r^2 d\Omega_{D-1}^2 + \fft{dr^2}{1 + \left(\ft{a}{r}\right)^{2(D-2)}}
\Big) - e^{\lambda_0\rho} d\tilde s_p^2 +
\sum_{i=p}^{n-1} e^{\lambda_i \rho}
dz_i^2\,,\cr
p\lambda_0 + \sum_{i=p}^{n-1} \lambda_i &=& \lambda\,,\qquad
p\lambda_0^2+ \sum_{i=p}^{n-1} \lambda_i^2 + \fft{\lambda^2}{D-2} =
4 (D-1)(D-2) a^{2(D-2)}\,,\cr
\rho &=& - \fft{1}{(D-2)a^{D-2}} {\rm arcsinh}\left(
\fft{a}{r}\right)^{D-2}\,,
\eea
where $p=2,3$ and the $d\tilde s_p^2$'s are given by
\bea
ds_2^2 &=&-du dv + \alpha \rho\, dv^2\,,\cr
ds_3^2 &=& -du dv + dw^2 - (\alpha \cos\beta\, \rho -
\ft18 \alpha^2\sin^2\beta\, \rho^2) dv^2 -
\alpha\sin\beta\, \rho\, dv dw\,.
\eea
In general the solution has
a naked curvature power-law singularity at $r=0$.
When all the $\lambda_i$ vanish, the solution reduces to either the
pp-wave for $p=2$ and the tachyon wave for $p=3$.

\section{Conclusions}

     In this paper, we construct the most general Ricci-flat metrics
in $(D+n)$ dimensions with the $\R^{1,n-1}\times SO(D)$ isometry.  We
find there are three classes of solutions.  The first class describes
the smooth Lorentzian wormholes that connect two asymptotic-flat
spacetimes, as well as its analytic extension which has a naked
curvature power-law singularity in the middle.  All the wormhole
metrics have one off-diagonal component in the $\R^{1,n-1}$ directions
so that the solutions have both mass and a linear momentum, which
propagates in one space direction. In the second class, the metric in
the $\R^{1,n-1}$ direction is diagonal, and hence the solution has
only the mass, with no linear momentum. The solution is asymptotic
Minkowskian, but with a naked curvature power-law singularity in the
middle. The third class describes a tachyon wave whose linear momentum
is larger than its mass.  The solution fits the general definition of
a pp-wave in that there exists a covariantly constant null vector.
However, the tachyon wave has its own distinct features.  The
world-volume for the tachyon wave is $\R^{1,2}$ instead of $\R^{1,1}$
for the usual pp-waves.  We verify, up to the cubic order, that the
Polynomial scalar invariants of the Riemann tensor vanish identically.
We expect that they all vanish identically, as in the case of the
pp-waves.  We also show in the appendix B that the tachyon waves admit
no Killing spinor except in $D=4$, in which case, it preserves half of
the supersymmetry.

      We also obtain $p$-brane solutions where the $\R^{1,n-1}$ part
of the spacetime lies in the world-volume of the $p$-branes.
Particularly interesting examples include M-branes and D3-brane, for
which AdS can arise in certain decoupling limits.  For the first
class, solutions become AdS wormholes that connect AdS$\times$Sphere
in one asymptotic region to a flat spacetime in the other.  The global
structure and traversability of the $SL(2,\R)$ wormholes were
discussed in detail in \cite{lumeiwh,adswh}.  (See also
\cite{aagc,aagccf}.)  We shall discuss these properties of the general
$GL(n,\R)$ wormholes in a future publication. For the third class we
obtain $p$-brane solutions with a tachyon wave propagating in the
world-volume.  In the decoupling limit of the corresponding M-branes
and D3-brane, the metric becomes a direct product of a sphere and an
AdS tachyon wave. We present the AdS tachyon wave for all dimensions
$D\ge 4$.  These solutions and the AdS wormholes provide interesting
backgrounds for the AdS/CFT correspondence.

\section*{Acknowledgement}

The research of H.L. and J.M. is supported in part by DOE grant
DE-FG03-95ER40917. Z.L.W. acknowledge support by grants
from the Chinese Academy of Sciences, a grant from 973 Program with
grant No: 2007CB815401 and grants from the NSF of China with Grant
No:10588503 and 10535060. We are grateful to Malcolm Perry for
useful discussions.

\appendix

\section{Properties of the matrix ${\cal C}$}

In this appendix, we study the properties of the $SL(n,\R)$
Lie-algebra valued integration constant matrix ${\cal C}$,
defined by
\be
{\cal M}^{-1} \dot {\cal M} \equiv {\cal C}\,.
\ee
(See section 2 for detail.)  We require all the elements of
${\cal C}$ to be real.  Under the $SL(n,\R)$ transformation
\be
d\vec z\rightarrow \Lambda^{-1} d\vec z\,,\qquad
{\cal M}\rightarrow \Lambda^{\rm T} {\cal M} \Lambda\,,
\ee
${\cal C}$ transforms as
\be
{\cal C} \rightarrow \Lambda^{-1} {\cal C} \Lambda\,.
\ee
We use the Borel transformation to fix the boundary condition
for ${\cal M}$ so that
\be
{\cal M}_\infty =\eta \equiv {\rm diag} (-1,1,\ldots,1)\,.
\label{M.diag}
\ee
 It follows that with this boundary condition $\eta
{\cal C}$ must be symmetric. The general ${\cal C}$ is
given by (\ref{gencmatrix}).

           There is a residual $SO(1,n-1)$ symmetry of $SL(n,\R)$
that preserves asymptotic ${\cal M}_\infty$.  We can use the
$SO(n-1)$ subgroup to simplify ${\cal C}$ so that we have
\be
{\cal C}=\begin{pmatrix}
-\alpha_0    & -\beta_1   & -\beta_2   & \cdots & -\beta_{n-1} \cr%
\beta_1     & \alpha_1    & 0      & \cdots & 0        \cr%
\beta_2     & 0      & \alpha_2    & \cdots & 0        \cr%
\vdots  & \vdots & \vdots & \ddots & \vdots   \cr%
\beta_{n-1} & 0      & 0      & \cdots & \alpha_{n-1}
\end{pmatrix}\,.\label{C.general}
\ee
Here $\alpha_0=\sum_{i=1}^{n-1}\alpha_i$ so that the matrix
is traceless.  Note that ${\cal C}$ can have real or complex
eigenvalues.  Since ${\cal C}$ is real, complex eigenvalues
should arise in pairs.  The eigenvectors, satisfying
\be%
{\cal C}^u_{~v}v^{(a)v}=\lambda^{(a)}v^{(a)u}\,,\qquad
{\cal C}^u_{~v}v^{\ast(a)v}=\lambda^{\ast(a)}v^{\ast(a)u}\,,
\label{C.eigen.vec}
\ee
form a linear space. The inner product of the eigenvectors is defined
by contracting indices with the Minkowski metric given in
Eq.(\ref{M.diag}). It follows that any {\it real} eigenvector
$\vec{v}$, associated with a real eigenvalue, can be {\it time-like,
space-like} or {\it null}, depending on whether $v^2\equiv
v_uv^u=-1,1$ or $0$.

      We now present a set of properties of ${\cal C}$,
which will be useful for its classification.
\begin{description}%
    \item[p.1]%
Using $\eta$ to raise or lower indices, we have ${\cal C}^u_{~v}=
{\cal C}_v^{~u}$.  It follows that for any eigenvector defined in
Eq.(\ref{C.eigen.vec}), we have
\be
v^{(a)}_u{\cal C}^u_{~v}={\cal C}_v^{~u}v^{(a)}_u={\cal C}_{vu}
v^{(a)u}=\lambda^{(a)}v^{(a)}_v~.
\ee
    \item[p.2]%
    Eigenvectors of different eigenvalues are orthogonal to each other.

    {\it Proof: By using {\bf p.1}, we have
\bea%
    v^{(b)}_u{\cal C}^u_{~v}v^{(a)v}&=&\lambda^{(a)}v^{(b)}_uv^{(a)u}=
\lambda^{(b)}v^{(b)}_vv^{(a)v}~,\cr
    \Longrightarrow~~~v^{(b)}_uv^{(a)u}&=&0\qquad{\rm if}\qquad
\lambda^{(a)}\neq\lambda^{(b)}~.
\eea}%
    As a result, the eigenvector of any real eigenvalue is orthogonal
    to both the real and imaginary part of the eigenvector of any
    complex eigenvalue.
    \item[p.3]%
    All the time-like and {\it linearly-independent} null
    vectors have non-vanishing inner product with each other.

    {\it Proof:\begin{itemize}%
         \item%
         For any two time-like vectors, one can always choose a
normal orthogonal basis such that they are in the form
\be%
\vec{v}^{(1)}=a\{1,0,\cdots,0\}\,;\qquad \vec{v}^{(2)}=
b\{\cosh\delta,\sinh\delta,0,\cdots,0\}~.
\ee%
         \item%
         For any two linearly independent null vectors,
we can always choose a normal orthogonal basis so that they
are in the form
\be%
         \vec{v}^{(1)}=a\{1,1,0,\cdots,0\}\,;\qquad
\vec{v}^{(2)}=b\{1,\cos\delta,\sin\delta,0,\cdots,0\}~.
\ee%
         \item%
        For a time like vector and a null
vector, we can always choose a normal orthogonal basis such that
they are in the form
\be%
\vec{v}^{(1)}=a(1,0,0\dots)\,;\qquad \vec{v}^{(2)}=b(1,1,0,\dots)~.
\ee%
It is clear that $\vec v^{(1)}$ and $\vec v^{(2)}$ have a non-vanishing
inner product in all three cases above.
         \end{itemize}}%
    \item[p.4]%
    The matrix ${\cal C}$ can have at most one pair of complex
    eigenvalues.

    {\it Proof : Suppose there exist two pairs of different complex
    eigenvalues $\lambda^{(1)},\lambda^{(1)\ast}$ and
$\lambda^{(2)},\lambda^{(2)\ast}$ ,
    and the corresponding eigenvectors $\vec{x}\pm i\vec{y}$ and
$\vec{u}\pm i\vec{v}$ .
    Because of {\bf p.2} , one has
\bea%
    \vec{x}^2+\vec{y}^2=\vec{u}^2+\vec{v}^2~;\label{app.cons.1}\\
    \vec{x}\cdot\vec{u}=\vec{x}\cdot\vec{v}=\vec{y}\cdot\vec{u}=
    \vec{y}\cdot\vec{v}=0~.\label{app.cons.2}
\eea%
    Eq.(\ref{app.cons.1}) implies that one in $\vec{x},\vec{y}$
    and one in $\vec{u},\vec{v}$ must be either time-like or null.
    However, this is in contradiction with {\bf p.3} and
    Eq.(\ref{app.cons.2}). So there can be at most one pair of
    complex eigenvalues.}

    \item[p.5]%
     Any null subspace can only have one null direction.

     {\it Proof:
     Suppose there are two linearly-independent null vector
      $\vec{v}^{(1)}$ and $\vec{v}^{(2)}$.
     Without loss of generality, we can choose
\be%
\vec{v}^{(1)}=a(1,1,0,0\dots)\,;\qquad
\vec{v}^{(2)}=b(1,\cos\theta,\sin\theta,0\dots)~,
\ee%
then $\ft{\vec{v}^{(1)}}{a}+\fft{\vec{v}^{(2)}}{b}$ is timelike
thus the space is timelike. Therefore, null subspace can has only
one null direction.}
\end{description}%
We can now use these properties to classify the matrix ${\cal C}$.

\bigskip

\bigskip
\noindent{\bf Class I:}
\bigskip

In this class, ${\cal C}$ has one and only pair of complex eigenvalues,
as is allowed by {\bf p.4}.  In this case, we should not diagonalise the
${\cal C}$ since we require that ${\cal C}$ to be a real matrix.

From the proof of {\bf p.4}, we see that the existence of a pair of
eigenvalues means the existence of a time-like or null vector.  Then
from {\bf p.2} and {\bf p.3}, the remaining real eigenvalues of ${\cal
C}$ must all have space-like eigenvectors. These space-like
eigenvectors can be used to partially diagonalise ${\cal C}$ . So if
the pair of complex eigenvalues is $p$-fold degenerate, we have
\be%
{\cal C}=\begin{pmatrix} N_{2p\times2p} \cr%
              & \lambda_{2p} \cr%
              &         & \ddots \cr%
              &         &        & \lambda_{n-1}
\end{pmatrix}~,
\ee%
with
\be%
N_{2p\times2p}=\begin{pmatrix}
-\alpha_0    & -\beta_1   & -\beta_2   & \cdots & -\beta_{2p-1} \cr%
\beta_1     & \alpha_1    & 0      & \cdots & 0        \cr
\beta_2     & 0      & \alpha_2    & \cdots & 0        \cr%
\vdots  & \vdots & \vdots & \ddots & \vdots   \cr%
\beta_{2p-1}& 0      & 0      & \cdots & \alpha_{2p-1}
\end{pmatrix}~.
\ee%
As we shall see presently, all $\beta_i$ are non-vanishing,
and
\be%
\alpha_i\neq \alpha_j\,,\qquad {\rm if}\quad i\neq j\qquad \hbox{for}
\qquad i,j=1,\cdots,2p-1~.
\ee%
Since $N_{2p\times2p}$ has a pair of $p$-fold degenerate complex
eigenvalues,
\bea%
&&Det(N_{2p\times2p}-\lambda I_{2p\times2p})=(a+{\rm i}\,b-\lambda)^p
(a-{\rm i}\, b-\lambda)^p\,,\cr
\Longrightarrow&&\beta_i=\sqrt{\displaystyle
\frac{\left[(\alpha_i-a)^2+b^2\right]^p
}{\prod_{j\neq i}(\alpha_i-\alpha_j)}}\,,\qquad i,j=1,\cdots,2p-1~;\cr
&&a=Tr(N_{2p\times2p})/(2p)~.
\eea%
It is easy to see that the constants $\beta_i$ cannot be all real
for $p\geq2$ . However for $p=1$ , we have
\be%
a=\frac{\alpha_1-\alpha_0}{2}\,,\qquad
\beta_1=\sqrt{(\alpha_1-a)^2+b^2}~.
\ee%
So $N_{2p\times2p}$ exists only for $p=1$ .  In this case, we can
use $O(1,1)$ symmetry to set $\alpha_0=\alpha_1$.  Thus the
${\cal C}$ in this case can all be simplified by
the $SO(1,n-1)$ symmetry to be
\be%
{\cal C}=\begin{pmatrix}
\alpha & -\beta \cr%
\beta   & \alpha \cr%
& & \lambda_2 \cr%
& & & \ddots \cr%
& & & & \lambda_{n-1}
\end{pmatrix}~,\label{C.case.II}
\ee%

\bigskip
\noindent{\bf Class II:}
\bigskip

All the eigenvalues are real and ${\cal C}$ has one time-like
eigenvector $\vec{x}_{(0)}$.  In fact, as we can see from {\bf p.3},
there can be no more than one time-like eigenvector in constructing an
orthogonal basis.  We build a normal orthogonal basis
$\vec{x}_{(\mu)}$ based on $\vec{x}_{(0)}$
\bea
x_{(\mu)}^{\rho}x_{(\nu)\rho}=\eta_{(\mu)(\nu)}.
\eea
Making an $SO(1,n-1)$ transformation as
\be
\Lambda_{\mu}^{~\rho}=x_{(\mu)}^{\rho},
\ee
then ${\cal C}$ can be reduced to
\bea {\cal C}\rightarrow\begin{pmatrix} \lambda
& 0 \\
0 &  {C_{(n-1)\times (n-1)}} \end{pmatrix}\,.
\eea
The Euclidean subspace associated with $C_{(n-1)\times (n-1)}$ can be
diagonalised by a further $SO(n-1)$ transformation. This is consistent
with that there could be no more time-like eigenvectors orthogonal to
$x_{(0)}$.  Thus in this class, ${\cal C}$ can be diagonalised by an
$SO(1,n-1)$ transformation, give by
\be
{\cal C} = {\rm diag}(\lambda_0, \lambda_1, \ldots,
\lambda_{n-1})\,.
\ee

\bigskip
\noindent{\bf Class III:}
\bigskip

       In this class, all the eigenvalues are real, but there is no
time-like eigenvector.  It follows that all the
eigenvectors must be space-like or null. We see from {\bf p.1} and
{\bf p.2} that all null eigenvectors must share the same eigenvalue
and thus they belong to the same eigenspace. In addition, with {\bf
p.5}, we conclude further that there is actually only one null
eigenvector.

If there is a space-like eigenvector $\vec{x}_{(n-1)}$, we can build
a normal orthogonal basis $\vec{x}_{(\mu)}$ based on
$\vec{x}_{(n-1)}$ and construct the corresponding $SO(1,n-1)$
transformation $\Lambda$ as in class II, which transforms ${\cal C}$
to
\be%
{\cal C}\rightarrow\begin{pmatrix}
{C_{(n-1)\times (n-1)}}& 0 \\
0 &  \lambda \end{pmatrix}~,\ee %
This procedure can be repeated until there is no more space-like
eigenvector in the remaining off-diagonal subspace. This leads to 
\be%
{\cal C}=\begin{pmatrix}
N_{p\times p} \cr%
& \lambda_p \cr%
& & \ddots \cr%
& & & \lambda_{n-1}
\end{pmatrix}~,
\ee%
 where $N_{p\times p}$ is a degenerate matrix with only one $p$-fold
eigenvalue $\lambda$ but only one eigenvector, which is null.
Obviously, we must have $p>1$; otherwise, 
the system would be reduced class II.

By $SO(p-1)$ transformations, we can take
\be%
N_{p\times p}=\begin{pmatrix}
-\alpha_0    & -\beta_1   & -\beta_2   & \cdots & -\beta_{p-1} \cr%
\beta_1     & \alpha_1    & 0      & \cdots & 0        \cr
\beta_2     & 0      & \alpha_2    & \cdots & 0        \cr%
\vdots  & \vdots & \vdots & \ddots & \vdots   \cr%
\beta_{p-1} & 0      & 0      & \cdots & \alpha_{p-1}
\end{pmatrix}~.\ee%
As we see presently, all the constants are non-vanishing, and
\be%
\alpha_i\neq \alpha_j\qquad{\rm if}\qquad i\neq j\qquad{\rm for}
\qquad i,j=1,\cdots,p-1~.
\ee%
Furthermore, Since $N_{p\times p}$ has a $p$-fold
degenerate eigenvalue, it follows that
\bea
&&{\rm Det}(N_{p\times p}-\lambda I_{p\times p})=
(\lambda_0-\lambda)^p\cr
\Longrightarrow&&\beta_i=
\sqrt{\displaystyle\frac{(\alpha_i-\lambda_0)^p}{\prod_{j\neq
i}(\alpha_i-\alpha_j)}}\,, \qquad i,j=1,\cdots,p-1~;\cr
&&\lambda_0=\frac{1}{p}~Tr(N_{p\times p})~.
\eea%
It can be easily seen that the $\beta_i$  cannot be all real for
$p\geq4$ . So $N_{p\times p}$ exists only for $p=2$ and $p=3$.
Indeed, both cases have one null eigenvector.

 Let us first look at the case with $p=2$.  We have
\be
\lambda_0=\frac{\alpha_1-\alpha_0}{2}\,,\qquad
\beta_1=|\alpha_1-\lambda_0|=\frac{|\alpha_0+\alpha_1|}{2}~.
\ee%
For $n=2$, the traceless condition implies that
\be
{\cal C}_2=\alpha \begin{pmatrix}
1 &-1 \cr
1 &-1 \end{pmatrix}\,,
\ee
For $n\ge 3$, we have
\be
{\cal C} = \begin{pmatrix}
{\cal C}_2 +\lambda_0 \oneone &   &    & \cr
                              &\lambda_2 & & \cr
                              &          & \ddots & \cr
                              & & & \lambda_{n-1}
\end{pmatrix}
\ee
Now let us consider $p=3$.  We have
\be
\lambda_0=\ft13 (\alpha_1 + \alpha_2 - \alpha_0)\,,\qquad
\beta_1 = \sqrt{\fft{(\alpha_1-\lambda_0)^3}{\alpha_1-\alpha_2}}\,,
\qquad \beta_2=\sqrt{\fft{(\alpha_2-\lambda_0)^3}{\alpha_2-\alpha_1}}
\,.
\ee
For $n=3$, the traceless condition implies that $\lambda_0=0$.
We can parameterise $\alpha_1=\alpha \sin^2\ft12\beta$
and $\alpha_2=-\alpha \cos^2\ft12\beta$. Thus we have
\be {\cal C}_3 =\a\begin{pmatrix} \cos\beta & -\sin^3\ft12\beta &
-\cos^3 \ft12\beta\cr \sin^3\ft12\beta & \sin^2\ft12\beta & 0 \cr
\cos^3\ft12\beta & 0 & -\cos^2\ft12\beta
\end{pmatrix}\,.
\ee
The ${\cal C}_3$ given in sections 4 and 5 is related to this
by an $SO(2)$ transformation.  For $n\ge 4$, we thus have
\be
{\cal C} = \begin{pmatrix}
{\cal C}_3 +\lambda_0 \oneone &   &    & \cr
                              &\lambda_3 & & \cr
                              &          & \ddots & \cr
                              & & & \lambda_{n-1}
\end{pmatrix}
\ee
Note that the matrices ${\cal C}_p$ have rank $p-1$ for $p=2,3$.
In both cases, all eigenvalues are zero, with just one eigenvector,
which is null.  Also the proportion factor $\alpha$ in
${\cal C}_2$ and ${\cal C}_3$ can be set to 1 by a further boost. We
keep it since it is related to the momentum charge for the
corresponding waves.

      Finally it is worth pointing out that the above classification
also applies for the $GL(n,\R)$ system, in which case, the traceless
condition for ${\cal C}$ is relaxed.

\section{Killing spinor analysis for the tachyon wave}

The tachyon wave solution in $D\ge 4$ dimensions
we obtained in section 4 can be rewritten as
\be
ds^2_D = -du dv + dw^2 - \left(\alpha \rho \cos\beta - \ft18 \alpha^2
\rho^2\sin^2\beta\right) dv^2 - \alpha \rho \sin\beta\, dv dw
+\sum_{i=1}^{D-3} dx^i dx^i\,,\label{appendwave}
\ee
where $\rho$ is given by
\bea
\rho = r^{5-D}\,,&& \hbox{for}\qquad D\ge 6\,,\cr
\rho = \log r\,, && \hbox{for}\qquad D = 5\,,\cr
\rho = r\,, && \hbox{for}\qquad D = 4\,,
\eea
with $r\equiv \sqrt{x^ix^i}$ is the radial coordinate of the
transverse space $dx^i dx^i$.  We choose the following
vielbein basis
\bea
e^{\hat +} &=& \ft12 dv\,,\qquad
e^{\hat -} = du + (\alpha \rho \cos\beta + \ft18 \alpha^2\rho^2
\sin\beta) dv\,,\cr
e^{\hat w} &=& dw - \ft12 \alpha \rho \sin\beta dv\,,\qquad
e^{\hat i} = dx^i\,,
\eea
then we have
\be
ds^2 = -2e^{\hat +}e^{\hat -} + (e^{\hat w})^2 +
e^{\hat i} e^{\hat i}\,.
\ee
Note that we use hat to denote tangent indices.
The non-vanishing spin connections are given by
\bea
\omega_{\hat w\hat +} &=& - \ft12(D-5)\alpha \sin\beta
\fft{x_i}{r^{D-3}} dx^i\,,\cr
\omega_{\hat i\hat +} &=& -(D-5)\alpha \fft{x_i}{r^{D-3}}
(\cos\beta dv + \ft12\sin\beta dw)\,,\cr
\omega_{\hat i\hat w} &=& -\ft14 (D-5)\alpha\sin\beta
\fft{x_i}{r^{D-3}} dv\,.
\eea
Note that for $D=5$, we will take $(D-5)\alpha\equiv\tilde\alpha$
to be non-vanishing.
Substituting these into the Killing spinor equation
\be
\delta \psi_{M} = D_{M} \epsilon =
\left(\del_M + \ft14 \omega_{M}^{AB} \Gamma_{AB}\right) \epsilon=0\,,
\ee
we have
\bea
\left(\del_i - \ft14(D-5)\alpha\sin\beta \fft{x_i}{r^{D-3}}
\Gamma^{\hat w}\Gamma^{\hat +}\right)\epsilon &=&0\,,\label{ks1}\\
\left(\del_v -\ft12(D-5)\alpha \fft{x_i}{r^{D-3}} [\cos\beta\,
\Gamma^{\hat i}\Gamma^{\hat +} +
\ft14 \sin\beta\, \Gamma^{\hat i}\Gamma^{\hat w}]\right)\epsilon
&=&0\,,\label{ks2}\\
\left(\del_w -\ft14 (D-5)\alpha \sin\beta\, \fft{x_i}{r^{D-3}}
\Gamma^{\hat i}\Gamma^{\hat +}\right)\epsilon &=& 0\,,\label{ks3}\\
\del_u\epsilon &=&0\,,\label{ks4}
\eea

        In $D=4$, for generic $\beta$ value, solution is $\epsilon=
\exp (-\ft18 \alpha v\sin\beta \Gamma^{\hat x} \Gamma^{\hat
w})\epsilon_0 $, where $\epsilon_0$ is a constant spinor with
$\Gamma^{\hat +} \epsilon_0=0$, and hence the solution preserves half
of the supersymmetry.  When $\beta=0$, the metric is flat and hence it
preserves full supersymmetry.  Indeed, the Killing spinor is given by
$\epsilon=(1 - \ft12 \alpha v \Gamma^{\hat x}\Gamma^{\hat
+})\epsilon_0$, where $\epsilon_0$ is an arbitrary constant spinor.

        For $D\ge 5$, the $\beta=0$ case gives rise to the vacuum
pp-wave solution.  The corresponding Killing spinors are constant
spinors $\epsilon_0$ subject to $\Gamma^{\hat +} \epsilon_0=0$.  For
generic $\beta\ne0$, the equations (\ref{ks1},\ref{ks4}) can be easily
solved, given
\be
\epsilon=\exp\left[-\ft14\alpha \sin\beta \fft{1}{r^{D-5}}
\Gamma^{\hat w}\Gamma^{\hat +}\right]\tilde \epsilon(v,w)=
\left(1 - \ft14\alpha \sin\beta \fft{1}{r^{D-5}}
\Gamma^{\hat w}\Gamma^{\hat +}\right)\tilde
\epsilon(v,w)\,.
\ee
Substituting this into (\ref{ks2}), we have
\be
\left(\del_w -\ft14(D-5)\alpha\sin\beta\,\fft{x_i}{r^{D-3}}
\Gamma^{\hat i}\Gamma^{\hat +}\right)\tilde \epsilon(v,w)=0\,.
\ee
Since this equation holds for arbitrary values of $x_i$, it follows
that $\Gamma^{\hat +} \tilde \epsilon (v,w)=0$ and
$\tilde \epsilon(v,w)=\tilde \epsilon(v)$.  The equation
(\ref{ks3}) then becomes
\be
\left (\del_v - \ft18 \alpha\sin\beta \fft{x_i}{r^{D-3}}
\Gamma^{\hat i} \Gamma^{\hat w}\right) \tilde\epsilon(u)=0\,.
\ee
It is clear that this equation has no solution for non-vanishing
$\beta$.

     It is also instructive to examine the integrability condition
\be
R_{\mu\nu ab} \Gamma^{ab}\,\epsilon=0\,.
\ee
The non-zero components of the curvature 2-form are
\bea
\Theta_{\hat w\hat +} &=& - \fft{(D-5)^2\alpha^2 \sin^2\beta}{
8r^{2(D-4)}} dv\wedge dw\,,\cr
\Theta_{\hat i\hat +} &=& (D-5)\alpha\cos\beta\,dv\wedge
d\left(\fft{x_i}{r^{D-3}}\right) +
\ft12(D-5)\alpha\sin\beta\, dw \wedge d\left(\fft{x_i}{r^{D-3}}
\right)\cr
&&+\ft18 (D-5)^2\alpha^2\sin^2\beta\, \fft{x_ix_j}{r^{2(D-3)}}
dv\wedge dx^j\,,\cr
\Theta_{\hat i\hat w} &=& \ft14 (D-5)\alpha \sin\beta
\wedge d\left(\fft{x_i}{r^{D-3}}\right)\,.
\eea
Thus for non-vanishing $\beta$, we have
\be
0=R_{vwab} \Gamma^{ab} \epsilon=-
\fft{(D-5)^2\alpha^2\sin^2\beta}{8 r^{2(D-4)}} \Gamma^{\hat w}
\Gamma^{\hat +} \epsilon
\qquad \Longrightarrow \qquad \Gamma^{\hat +} \epsilon=0\,.
\ee
The remaining conditions become
\be
R_{viab} \Gamma^{ab} \epsilon =\fft{(D-5)\alpha \sin\beta}{r^{D-1}}
\left(r^2 \Gamma^{\hat i}\Gamma^{\hat w} - (D-3) x_i x_j
\Gamma^{\hat j} \Gamma^{\hat w}\right)\epsilon=0\,.
\ee
Contract the equation with $x^i$, we have
\be
(D-4)x_i \Gamma^{\hat i} \Gamma^{\hat w} \epsilon=0\,.
\ee
It has no non-trivial solution except in four dimensions.

     Thus we have demonstrated that the general tachyon wave solutions
(\ref{appendwave}), with $\beta\ne 0$, admit no Killing spinors except
in $D=4$, in which case, it preserves half of the supersymmetry.

\end{document}